\documentclass[aps,prl,twocolumn,amsmath,amssymb,showpacs, superscriptaddress,notitlepage,longbibliography,10pt]{revtex4-1}

\usepackage[colorlinks=true,linkcolor=blue,anchorcolor=red,citecolor=blue, urlcolor=blue]{hyperref}
\usepackage{bm}
\usepackage{graphicx}
\usepackage{color}
\usepackage{multirow}
\usepackage{extarrows}
\usepackage{cases}

\begin{document}
\title{Negative Magnetoresistance without Chiral Anomaly in Topological Insulators}
\author{Xin Dai}
\affiliation{Institute for Advanced Study, Tsinghua University, Beijing 100084, China}
\author{Z. Z. Du}
\affiliation{Institute for Quantum Science and Engineering and Department of Physics, South University of Science and Technology of China, Shenzhen 518055, China}
\affiliation{School of Physics, Southeast University, Nanjing 211189, China}
\affiliation{Shenzhen Key Laboratory of Quantum Science and Engineering, Shenzhen 518055, China}

\author{Hai-Zhou Lu}
\email{Corresponding author:\\luhz@sustc.edu.cn}
\affiliation{Institute for Quantum Science and Engineering and Department of Physics, South University of Science and Technology of China, Shenzhen 518055, China}
\affiliation{Shenzhen Key Laboratory of Quantum Science and Engineering, Shenzhen 518055, China}

\date{\today}
\begin{abstract}
An intriguing phenomenon in topological semimetals and topological insulators is the negative magnetoresistance (MR) observed when a magnetic field is applied along the current direction.
A prevailing understanding to the negative MR in topological semimetals is the chiral anomaly, which, however, is not well defined in topological insulators.
We calculate the MR of a three-dimensional topological insulator, by using the semiclassical equations of motion, in which the Berry curvature explicitly induces an anomalous velocity and orbital moment.
Our theoretical results are in quantitative agreement with the experiments. The negative MR is not sensitive to temperature and increases as the Fermi energy approaches the band edge. The orbital moment and g factors also play important roles in the negative MR. Our results give a reasonable explanation to the negative MR in 3D topological insulators and will be helpful in understanding the anomalous quantum transport in topological states of matter.
\end{abstract}

\maketitle

\begin{figure}[htpb]
\centering
\includegraphics[width=0.9\columnwidth]{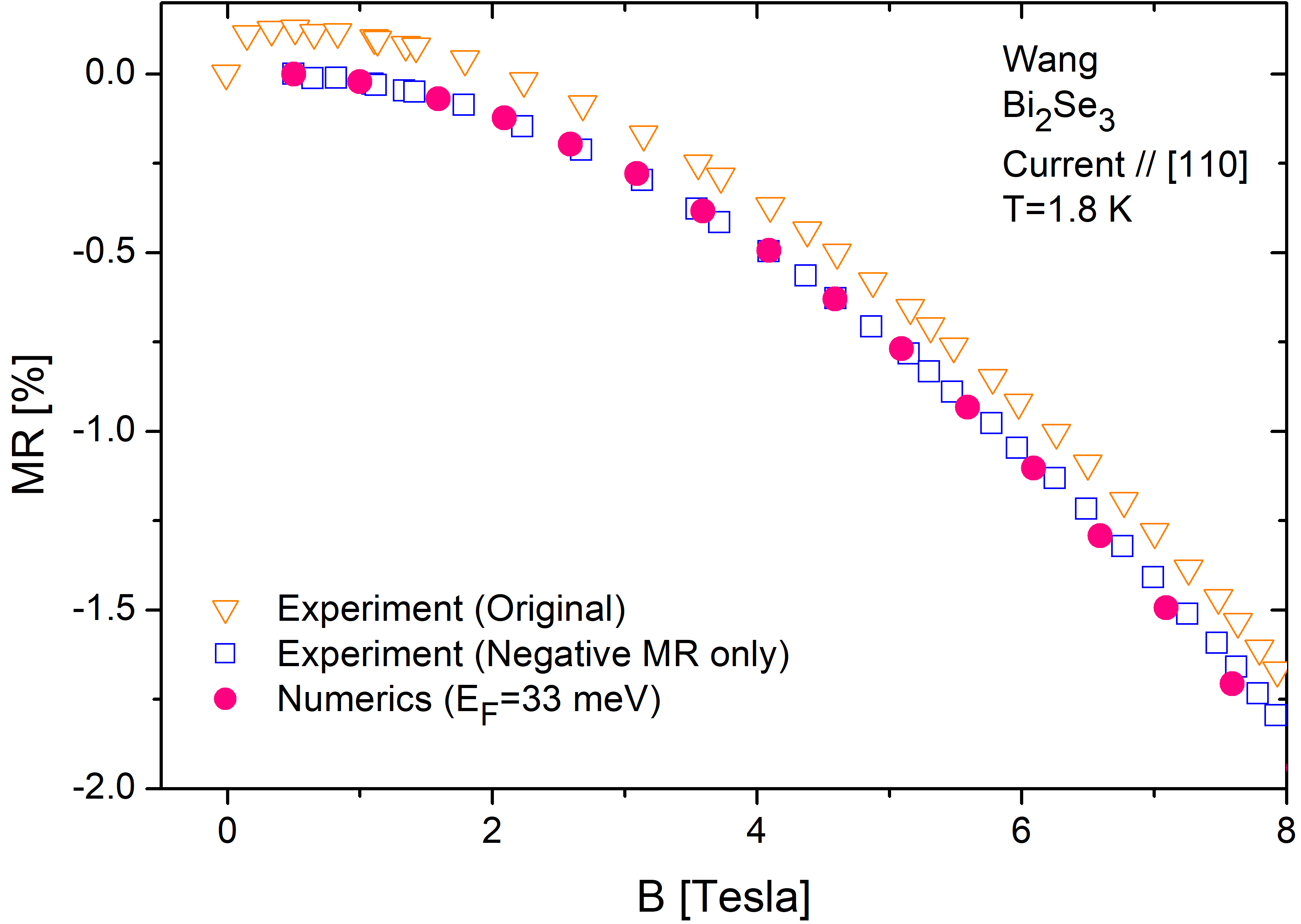}
\includegraphics[width=0.9\columnwidth]{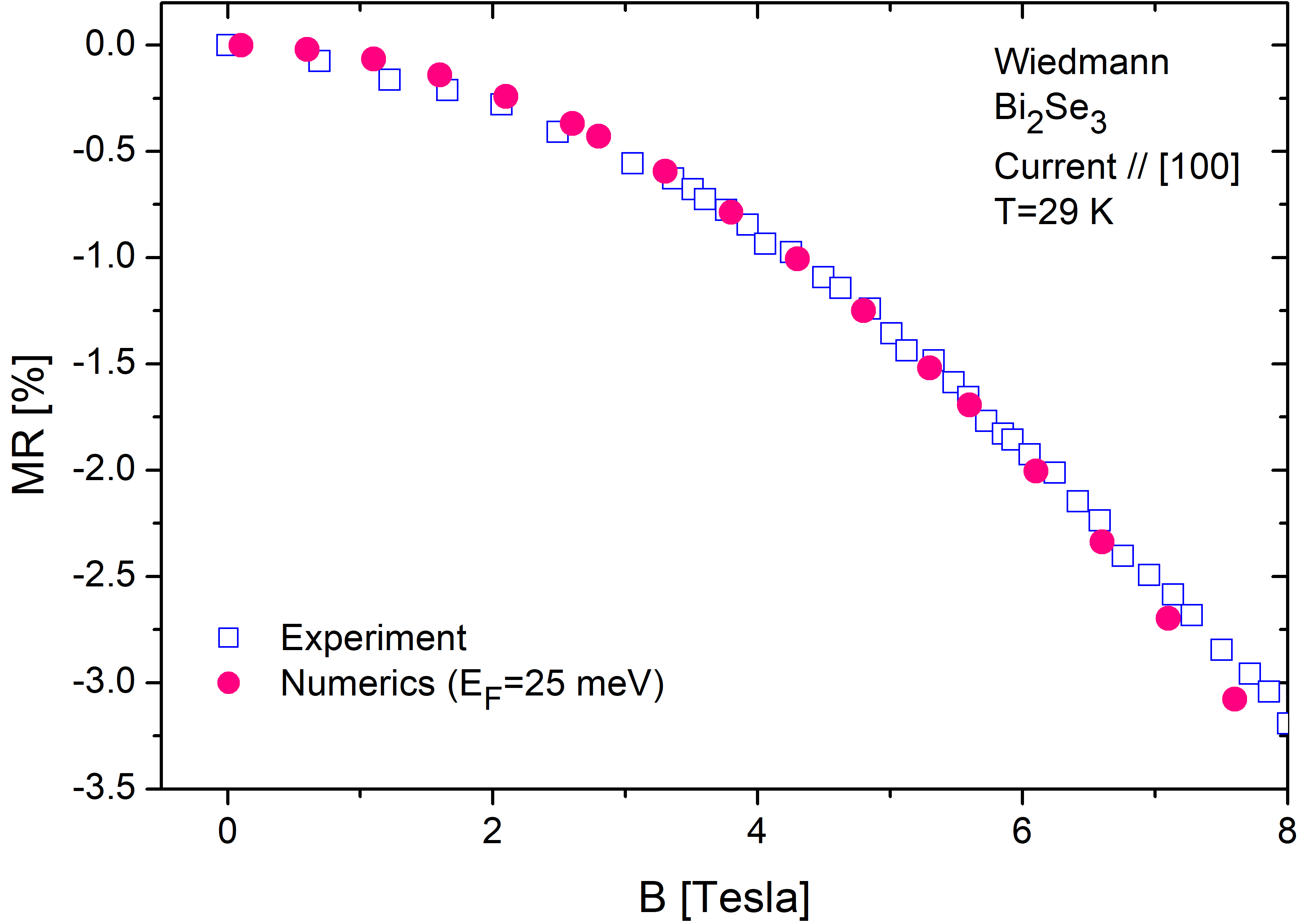}
\includegraphics[width=0.9\columnwidth]{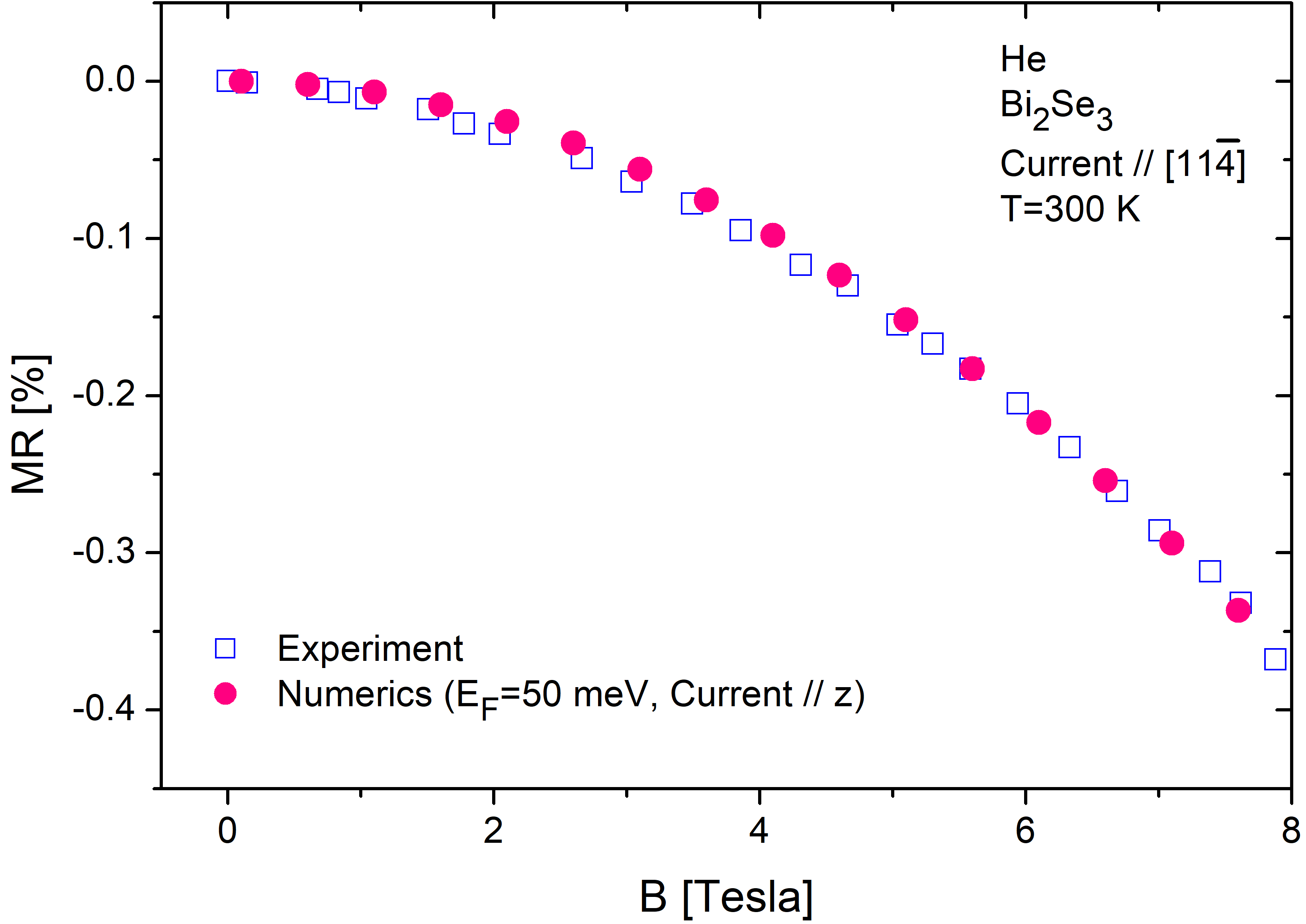}
\caption{The comparison between the calculated negative MR and the experiments at three typical temperatures and current directions \cite{Wang12nr,Wiedmann16prb,He13apl}. In the calculations, the corrections from both the Berry curvature $\mathbf{\Omega}$ and orbital moment $\mathbf{m}$ have been taken into account. The Fermi energy $E_F$ (measured from the bottom of the conduction bands) is a tuning parameter in the numerics. All values of $E_F$ fall in a reasonable range. The current direction and temperature are from the experiments. $x,y,z$ in the model \eqref{full_hamiltonian} correspond to [100], [010], and [001] crystallographic directions, respectively. The numerically calculated MR along the $z$ axis is used to approach the experimental MR along the $[11\bar{4}]$ direction by He \emph{et al.}, because the projection of $[11\bar{4}]$ on the $z$ axis is over 94\%.
Other model parameters are from the $k\cdot p$ calculations \cite{Nechaev16prb} and experiments \cite{Wolos16prb}, $M_0$=-0.169 eV, $M_z$=3.351 eV$\mbox{\AA}^2$, $M_{\perp}$=29.36 eV $\mbox{\AA}^2$, $V_{\perp}$=2.512 eV$\mbox{\AA}$, $V_n$=1.853 eV$\mbox{\AA}$, $C_0$=0.048 eV, $C_z$=1.409 eV $\mbox{\AA}^2$, $C_{\perp}$=13.9 eV$\mbox{\AA}^2$, the g factors $g_{z}^v=g_{z}^c=30$ and $g_{p}^v=g_{p}^c=-20$.}
\label{Fig:comparison}
\end{figure}

{\color{red}\emph{Introduction}} -
Recently discovered topological semimetals are characterized by a negative magnetoresistance (MR) \cite{Kim13prl,Kim14prb,Li16np,ZhangCL16nc,HuangXC15prx,Xiong15sci,LiCZ15nc,ZhangC17nc,LiH16nc,Arnold16nc,YangXJ15arXiv,YangXJ15arXiv-NbAs,WangHC16prb}, which is rare in nonmagnetic materials.
The negative MR is widely believed to be a signature showing that a topological semimetal can host the chiral anomaly, that is, the conservation of chiral current is violated as a result of the quantization \cite{Adler69pr,Bell69Jackiw,Nielsen81npb}.
However, in other systems where the chiral anomaly is not well defined, e.g., in topological insulators, a negative MR has also been observed and has created great confusion \cite{Wang12nr,He13apl,Wiedmann16prb,Wang15ns,Breunig17nc,Assaf17arXiv}.
In this Letter,
we present a quantitative study on the MR of 3D topological insulators.
Using the semiclassical Boltzmann formalism, we explicitly take into account the correction to the conductivity from the anomalous velocity induced by the Berry curvature and orbital moment of the bulk states. By using the parameters for Bi$_2$Se$_3$, we find that the MR can be negative when the magnetic field is applied parallel with the current and in a quantitative agreement with the experiments (see Fig. \ref{Fig:comparison}).
Consistent with the experiments, the negative MR is not sensitive to temperature, as expected by its semiclassical nature. The negative MR depends on the Fermi energy, and its magnitude increases when approaching the band edge. We also find that the MR depends on the signs of g factors, and may provide an approach to measure the g factors for these materials. Our results may give a reasonable explanation to the experimentally observed negative MR in 3D topological insulators, and will be helpful for understanding the anomalous quantum transport in topological states of matter.\\

{\color{red}\emph{Anomalous velocity}} -
First, we illustrate that the anomalous velocity induced by the Berry
curvature and its derivative orbital moment is the reason behind the negative MR. In the experiments of topological insulators, the negative MR can survive above $T=100$ K~\cite{Wiedmann16prb},
so quantum interference mechanisms, such as the weak localization effect, can be excluded.
Moreover, because of the poor mobility in the topological insulators Bi$_2$Se$_3$ and Bi$_2$Te$_3$ \cite{Culcer12pe}, the Landau levels cannot
be well formed up to 6 T in the experiments.
In this semiclassical regime, the electronic transport can be described by the equations of motion \cite{Sundaram99prb}
\begin{eqnarray}\label{EOM-maintext}
\dot{\mathbf{r}}=\frac{1}{\hbar}\nabla_{\mathbf{k}}\widetilde{\varepsilon}_{\mathbf{k}}
-\dot{\mathbf{k}}\times \mathbf{\Omega}_{\mathbf{k}}, \ \ \ \
\dot{\mathbf{k}}=-\frac{e}{\hbar}(\mathbf{E}+\dot{\mathbf{r}}\times\mathbf{B}),
\end{eqnarray}
where both the position $\mathbf{r}$
and wave vector $\mathbf{k}$ appear simultaneously,
$\dot{\mathbf{r}}$ and $\dot{\mathbf{k}}$ are their time derivatives, $-e$ is the electron charge,
and $\mathbf{E}$ and $\mathbf{B}$ are external electric and magnetic fields,
respectively.  $\widetilde{\varepsilon}_{\mathbf{k}}=\varepsilon_{\mathbf{k}}-\mathbf{m} \cdot \mathbf{B}$,
$\varepsilon_{\mathbf{k}}$ is the band dispersion, $\mathbf{m}$ is the orbital moment
induced by the semiclassical self-rotation of the Bloch wave packet, and $\mathbf{\Omega}_{\mathbf{k}}$ is the Berry curvature \cite{Xiao10rmp}.

In the linear-response limit {\color{blue}($\mathbf{E}=0$)}, Eq.~\eqref{EOM-maintext} yields an effective velocity
\begin{eqnarray}\label{Eq:velocity}
&&\dot{\mathbf{r}}= [\widetilde{\mathbf{v}}_{\mathbf{k}}
+(e/\hbar)\mathbf{B}(\widetilde{\mathbf{v}}_{\mathbf{k}}\cdot\mathbf{\Omega}_{\mathbf{k}})]/D_{\mathbf{k}},
\end{eqnarray}
where $D_{\mathbf{k}}^{-1}$ is the correction to the density of states, and
\begin{equation}\label{v_and_d}
\begin{split}
\widetilde{\mathbf{v}}_{\mathbf{k}}=\mathbf{v}_{\mathbf{k}}-\frac{1}{\hbar}
\nabla_\mathbf{k} (\mathbf{m}_{\mathbf{k}}\cdot\mathbf{B}),\ \ \
D_{\mathbf{k}}=1+\frac{e}{\hbar}\mathbf{B}\cdot\mathbf{\Omega}_{\mathbf{k}}.
\end{split}
\end{equation}
Because of the Berry curvature, the velocity develops an anomalous term that is proportional to $\mathbf{B}$. Note that the conductivity is the current-current (velocity-velocity) correlation \cite{Mahan1990},
thus the presence of the anomalous velocity is expected to generate an extra conductivity that grows with the magnetic field. In other words, the Berry curvature and its derivative orbital moment may induce a negative MR.
It has been implied
that the negative MR in topological semimetals is related to the Berry curvature \cite{Son13prb,Yip15arXiv,Lu17fop}, which diverges near the Weyl nodes and can make a prominent contribution.
The concern is whether this mechanism is large enough in topological insulators as those observed in the experiments, where the relative MR
can exceed -1\% in a parallel magnetic field of several T \cite{Wang12nr,He13apl,Wang15ns,Wiedmann16prb,Breunig17nc,Assaf17arXiv}.
Later, we will use a realistic model of topological
insulator to show that the Berry curvature can lead to a negative MR comparable with the experiments.\\

\begin{figure}[htpb]
\centering
\includegraphics[width=0.9\columnwidth]{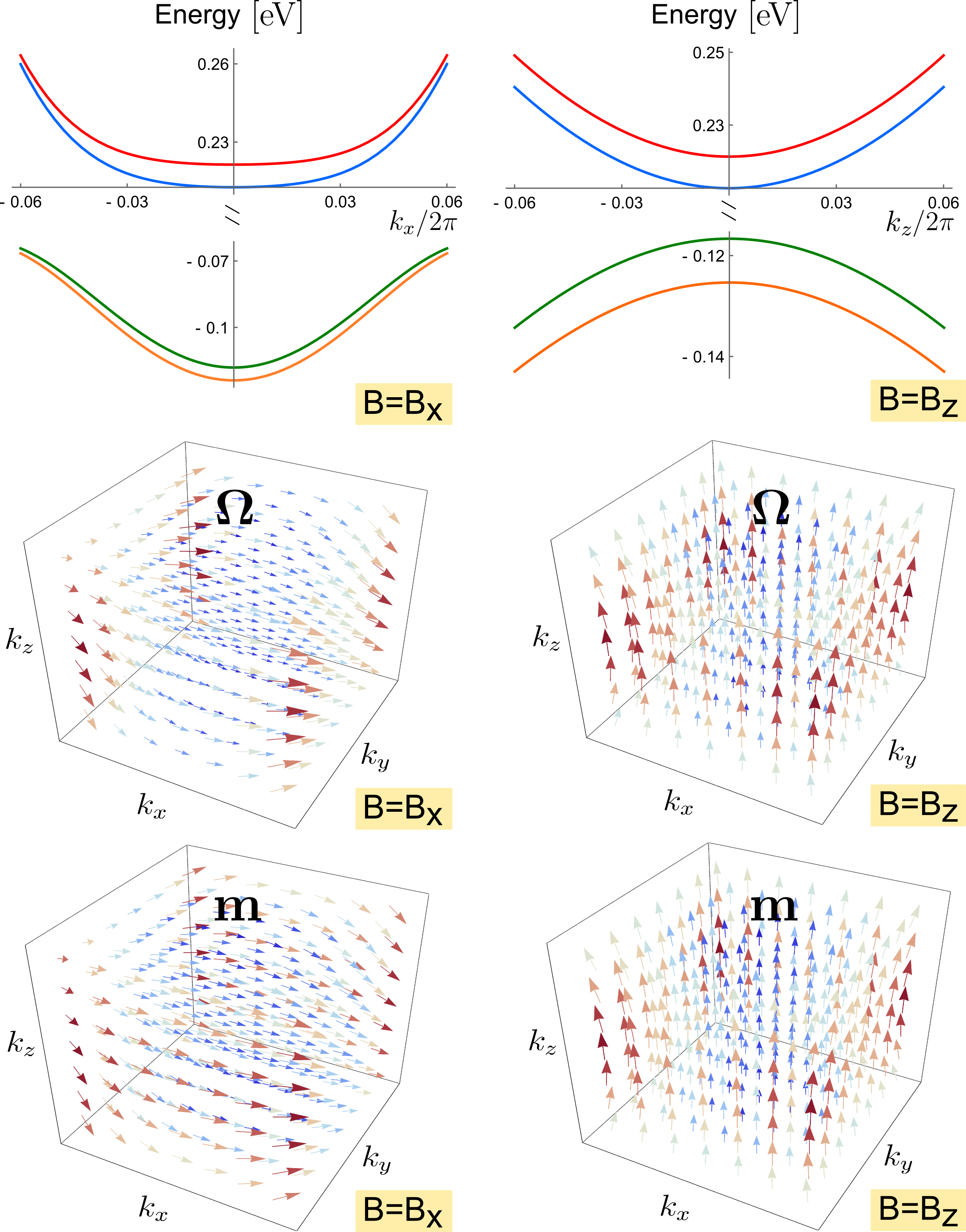}
\caption{The energy dispersion (top left $k_y$=$k_z$=0, top right $k_x$=$k_y$=0) of the $\mathbf{k}\cdot\mathbf{p}$ model, and the vector plots of the Berry curvature $\mathbf{\Omega}$ (middle) and orbital moment $\mathbf{m}$ (bottom) for the lower conduction band (blue curve).
The magnetic field is 5 T. The parameters are the same as those in Fig. \ref{Fig:comparison}.
}\label{Fig:Berry}
\end{figure}

{\color{red}\emph{Model and conductivity formula}} -
In the experiments, the negative MR occurs
in bulk samples which are at least several tens of nanometers
thick and the major carriers are from the 3D bulk states.
The 2D surface-state carriers can be neglected because 2D to 3D is like 0 to infinity.
A well accepted $k\cdot p$ Hamiltonian for the bulk states of 3D topological insulators
is \cite{Zhang09np,Nechaev16prb}
\begin{equation}\label{full_hamiltonian}
H_0=C_{\mathbf k}+
\begin{pmatrix}
M_{\mathbf k} & 0 & i V_n k_z & -i V_{\perp} k_-\\
0 & M_{\mathbf k} & i V_{\perp} k_+ & i V_n k_z\\
-i V_n k_z & -i V_{\perp} k_- & -M_{\mathbf{k}} & 0\\
i V_{\perp} k_+ & -i V_n k_z & 0 & -M_{\mathbf{k}}
\end{pmatrix},
\end{equation}
where $M_{\mathbf{k}}=M_0+M_{\perp}(k_x^2+k_y^2)+M_z k_z^2$, $C_{\mathbf{k}}=C_0+C_{\perp}(k_x^2+k_y^2)+C_z k_z^2$,
$M_i$, $V_i$, and $C_i$ are model parameters.
The model describes a 3D strong topological
insulator when $M_0M_\perp<0$ and $M_0M_z<0$ \cite{Shen12book}.
The model has four energy bands $\varepsilon_n(\mathbf{k})$ near the $\Gamma$ point, two conduction bands and two valence bands (see Fig \ref{Fig:Berry}).
We will assume that the Fermi level crosses only the two conduction bands.
In systems with both time- and centrosymmetric symmetries,
the Berry curvature vanishes at every $\mathbf{k}$ point in the Brillouin zone,
which is the case in pristine 3D topological insulators.
In the presence of the magnetic field,
a nonzero distribution of the Berry curvature can be induced by the Zeeman effect that breaks time reversal symmetry. The Zeeman Hamiltonian reads
\begin{eqnarray}\label{zeeman}
    H_Z=\frac{\mu_B}{2}\left(
    \begin{array}{cccc}
        g_{z}^{v}B_z  & g_{p}^vB_- & 0 & 0\\
        g_{p}^v B_{+} & -g_{z}^v B_z & 0 & 0\\
        0 & 0 &  g_{z}^c B_z & g_{p}^c B_- \\
        0 & 0 & g_{p}^c B_+ & -g_{z}^c B_z
    \end{array}
    \right),
\end{eqnarray}
where $\mu_B$ is the Bohr magneton and $g_{v/c,z/p}$ are Land\'{e} g factors
for valence and conduction bands along the $z$ direction and in the $x-y$ plane, respectively. 
The Zeeman energy can induce an anisotropy of the Fermi surface. But without $\mathbf{\Omega}$ and $\mathbf{m}$, the anisotropy alone does not contribute to the magnetoresistance \cite{Pal10prb}.

In our calculation, the relative MR is defined as
\begin{eqnarray}\label{MR-parallel}
\mathrm{MR}_{\mu}(B_\mu) =\frac{1/\sigma^{\mu\mu}(B_\mu)-1/\sigma^{\mu\mu}(0)}{1/\sigma^{\mu\mu}(0)}.
\end{eqnarray}
In the semiclassical Boltzmann formalism, the longitudinal conductivity $\sigma^{\mu\mu}$ is contributed by all the bands crossing the Fermi energy, and for band $n$ \cite{Yip15arXiv} (Sec. S1 of the Supplemental Material~\cite{Supp})
\begin{equation}
\sigma^{\mu\mu}\label{conduction_tensor}
=\int\frac{d^{3}\mathbf{k}}{(2\pi)^{3}}
\frac{e^{2}\tau}{D_{\mathbf k}}
\left(\widetilde{v}^{\mu}_{\mathbf{k}}
+\frac{e}{\hbar}B^{\mu}\widetilde{v}^{\nu}_{\mathbf{k}}\Omega^{\nu}_{\mathbf{k}}\right)^2
\left(-\frac{\partial\widetilde{f}_0}{\partial\widetilde{\varepsilon}}\right),
\end{equation}
where $n$ is suppressed for simplicity, $D_{\mathbf{k}}$ and $\widetilde{v}^{\mu}_{\mathbf{k}}$ are
given by Eq.~\eqref{v_and_d}, 
$\widetilde{f}_0$ is the equilibrium Fermi distribution,
the transport time $\tau$ is assumed to be a constant in
the semiclassical limit \cite{Burkov14prl-chiral}.
For the $n$th band of the Hamiltonian $H$,
the $\xi$ component of the Berry curvature
vector can be found as
$\Omega_{n\mathbf{k}}^{\xi}=\Omega_{n\mathbf{k}}^{\mu\nu}\varepsilon_{\mu\nu\xi}$,
where $\xi,\mu,\nu$ stand for
$x,y,z$, $\varepsilon_{\mu\nu\xi}$ is the Levi-Civita antisymmetric tensor, and
\begin{eqnarray} \label{berry}
\Omega_{n\mathbf{k}}^{\mu\nu}  = -2\sum_{n'\neq n}
\frac{\mathrm{Im}\langle n|
\partial H/\partial k_{\mu} |n'\rangle \langle n'|
  \partial H/\partial k_{\nu} |n\rangle}{(\varepsilon_n-\varepsilon_n')^2},
\end{eqnarray}
where $H=H_0+H_Z$. The orbital moment $\mathbf{m}$ can be found as
\begin{equation}\label{orbital}
m_{n\mathbf{k}}^{\mu\nu}  = -\frac{e}{\hbar}\sum_{n'\neq n}
\frac{\mathrm{Im}\langle n|
\partial H/\partial k_{\mu}  |n'\rangle \langle n'|
  \partial H /\partial k_{\nu} |n\rangle}
{\varepsilon_n-\varepsilon_n'}.
\end{equation}
Figure \ref{Fig:Berry} (and Fig. S1 of \cite{Supp}) show that the Zeeman energy can induce a finite distribution of $\mathbf{\Omega}$ and $\mathbf{m}$, which should be zero without the Zeeman energy.

{\color{red}\emph{Comparison with negative MR in experiments}} -
Figure~\ref{Fig:comparison} shows that the numerically calculated relative MRs (see the numerical scheme in Sec. S3 of Ref.~\cite{Supp})
in parallel magnetic fields are negative and decrease
monotonically with the magnetic field.
They can be fitted by $-B^2$ at small magnetic fields and conform to the 
Onsager reciprocity MR$(B)$=MR$(-B)$. To justify our numerical scheme, we also use a tight-binding model \cite{Mao11prb} to perform the calculation (Sec. S4 of the Supplemental Material \cite{Supp}). 
The $k\cdot p$ and tight-binding models give the same results 
at weak magnetic fields (Fig. S2 of Ref.~\cite{Supp}).
Figure \ref{Fig:comparison} shows a good agreement on the negative MR between the typical experiments and our numerical calculations. In our calculations, the Fermi energy $E_F$ is a tuning parameter. All values of $E_F$ fall in a reasonable range. The current direction and temperature are from the experiments and the model parameters are from the $k\cdot p$ calculations \cite{Nechaev16prb} and experiments \cite{Wolos16prb}. In the experiment by Wang \emph{et al}. \cite{Wang12nr}, 
the temperature is 1.8 K, so the original data (orange triangles) has a positive MR near zero field due to the weak antilocalization \cite{Checkelsky09prl,Chen10prl,Wang11prb,He11prl}. The competition between the weak antilocalization and the negative MR leads to a turning point at around 0.5 T. 
Thus, the comparison starts at 0.5 T, as shown by the pink and blue scatters. 
By contrast, in the experiments by Wiedmann \emph{et al}. \cite{Wiedmann16prb} 
and He \emph{et al}. \cite{He13apl}, the temperatures are 29 and 300 K, respectively, far above the critical temperature (about 10 K in Bi$_2$Se$_3$) of the weak antilocalization effect \cite{Checkelsky09prl,Chen10prl,Wang11prb,He11prl}. Therefore, there are only negative MR and the comparisons start from 0 T.  

\begin{figure}[htpb]
\centering
\includegraphics[width=0.9\columnwidth]{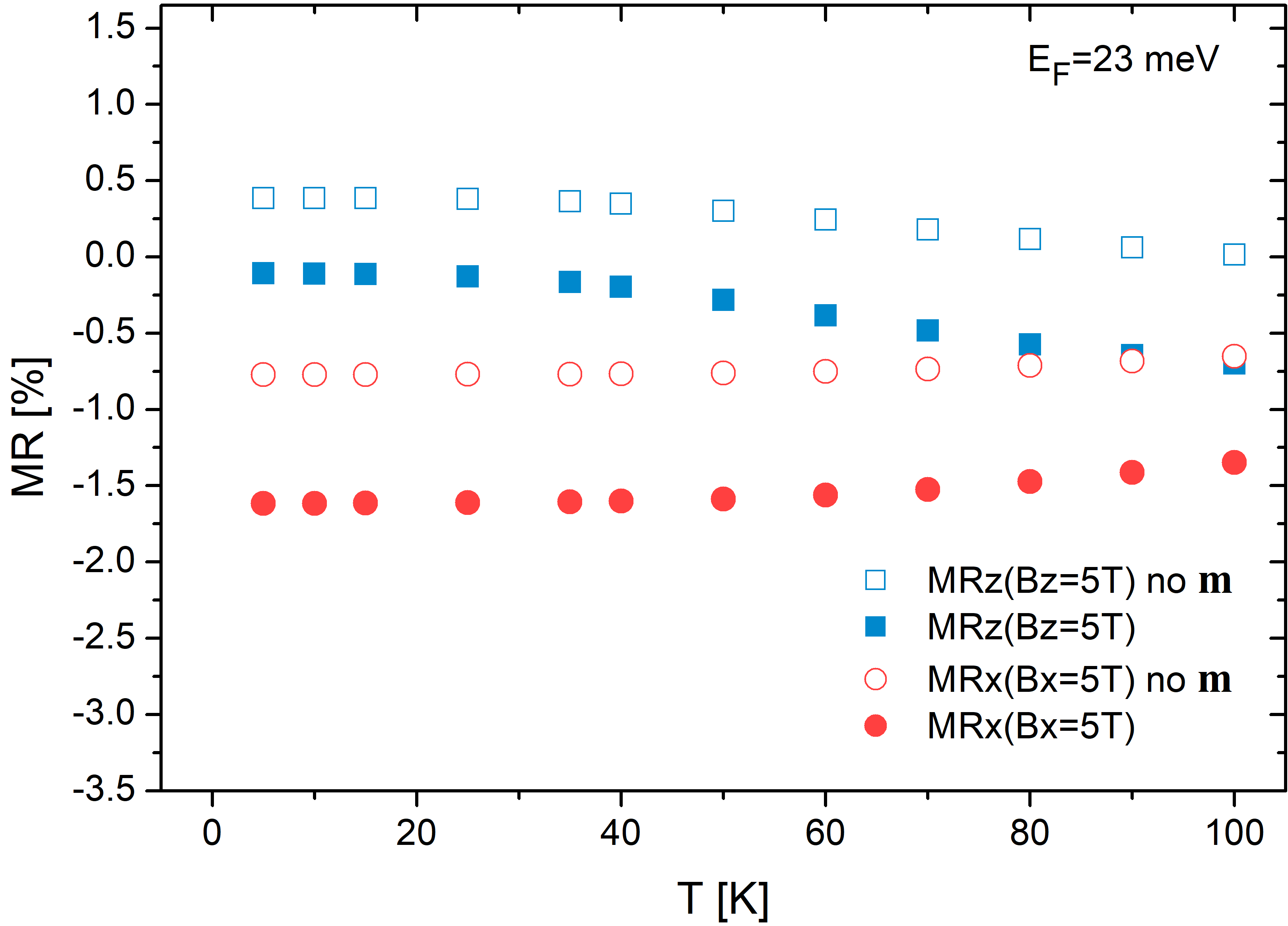}
\includegraphics[width=0.9\columnwidth]{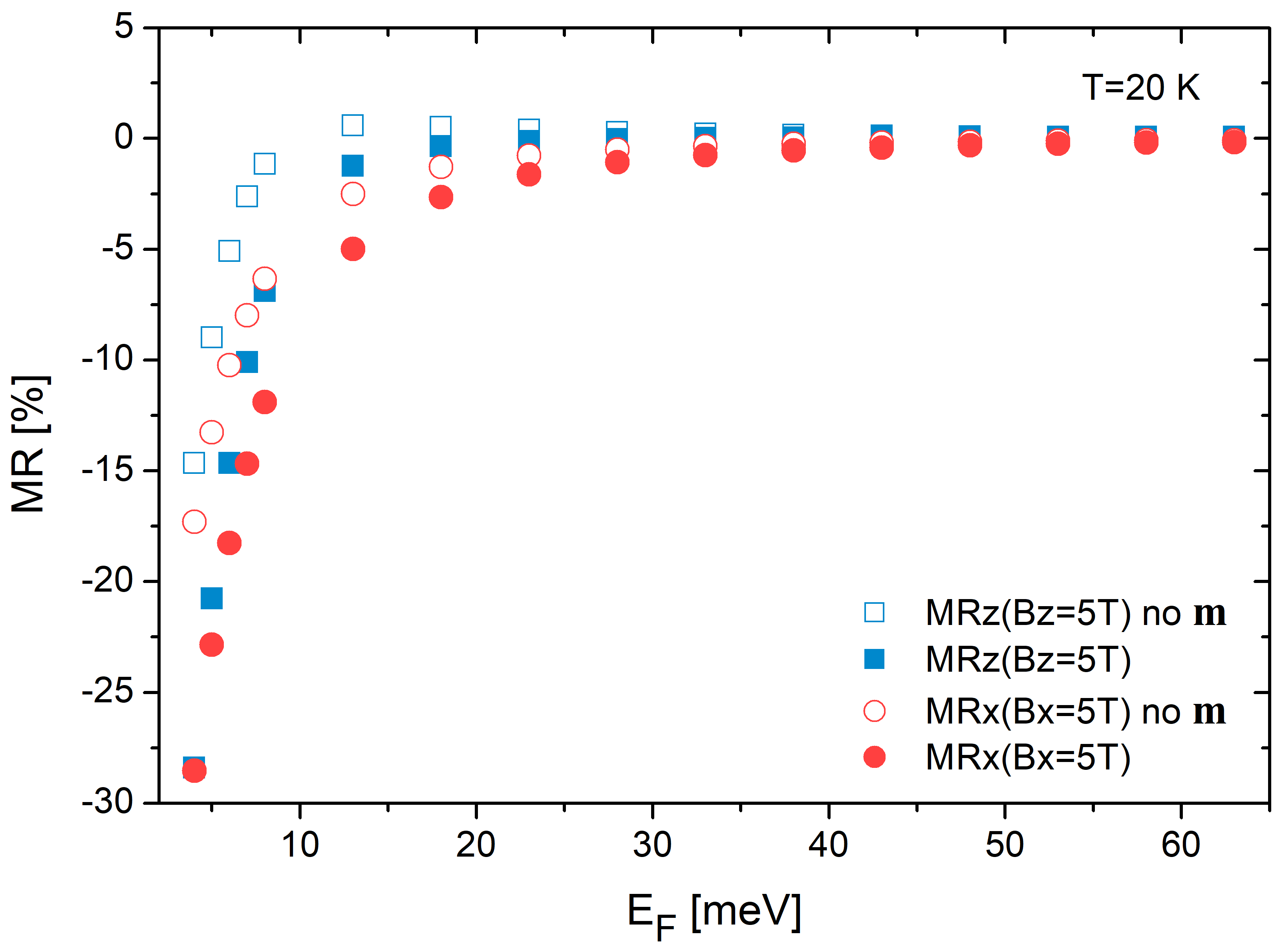}
\caption{The numerical results for MR$_{x}$($B_x$=5\;T) and MR$_{z}$($B_z$=5\;T) as functions of the Fermi energy $E_F$ and temperature $T$, in the presence (solid scatters) and absence (hollow scatters) of the orbital moment $\mathbf{m}$. The parameters are the same as those in Fig. \ref{Fig:comparison}. $E_F$ is measured from the bottom of the conduction bands. There is no qualitative change if using the parameters from Ref.~\cite{Liu10prb}. }\label{Fig:dependence}
\end{figure}

{\color{red}\emph{Temperature and Fermi energy dependences}} -
Now we show more detailed behaviors of the calculated negative MR.\
Figure~\ref{Fig:dependence} shows that
the negative MR does not change much with temperature.
This is consistent with the experiments, showing the semiclassical nature of the negative MR. Figure~\ref{Fig:dependence} also shows that the negative MR becomes enhanced as
the Fermi level approaches the band bottom. The enhancement of the negative MR near the band edge
can be understood using Fig.~\ref{Fig:Berry}, which shows that
the Zeeman splitting of the conduction bands is maximized at the $\Gamma$ point, about several meV in a magnetic field of 5 T. The negative MR is contributed by the Berry curvature from the
two conduction bands. In the absence of the Zeeman splitting,
both the Berry curvature and orbital moment vanish due to time-reversal and inversion symmetries. The Zeeman effect can break time-reversal symmetry and induce a finite distribution of the Berry curvature for the conduction bands. Therefore, the MRs increase with the magnitudes of Zeeman splitting and get enhanced near the band edge.

{\color{red}\emph{Roles of g factors}} -
We find that the signs of $g$-factors in the Zeeman coupling determine
the signs of MRs {\it qualitatively}.
Moreover, according to Eqs.~\eqref{berry} and \eqref{orbital},
the signs of $\bm \Omega$ and $\bf m$
are controlled by the signs of $g$-factors,
in particular $g_{z,p}^c$ when the
Fermi level crosses the conduction bands.
As pointed out earlier, the Zeeman splitting
is the largest at the $\Gamma$ point,
therefore the sign of the MR is determined by
the lower band with larger Fermi surface and the signs of the $g$-factors.
We list the relation between the signs of the
MRs and g factors in Tables. \ref{tab:g-Bz} and \ref{tab:g-Bx}.
The resulting MRs differs {\it qualitatively}
with different signs of $g_{z,p}^c$. Note that $g^v_{z,p}$ are irrelevant since we have assumed that the Fermi level crosses only the conduction bands.
In the experiment, the techniques used so far for topological insulators,
for example, electron spin resonance and quantum oscillations,
cannot determine the signs of $g$-factors but only their absolute values~\cite{Wolos16prb,Kohler75pssb}.
\color{black}
Transport measurements can determine
the sign of the $g$-factor only in specific setups \cite{Srinivasan16prbrc}.
Our theoretical calculation therefore provides a clue to evaluate the sign of $g$-factors.

\begin{table}[tb]
\caption{The relation between signs ($\pm$) of the MRs and g factors in the $z$-direction magnetic field.  }
\label{tab:g-Bz}%
\begin{ruledtabular}
\begin{tabular}{ccc}
$g_{z}^v$ & $g_{z}^c$ & MR$_{z}(B_z)$\\
\hline
+ & +   &  -
\\
- & +   & -
\\
- & -   & +
\\
+  & -  & +
\tabularnewline
\end{tabular}
\end{ruledtabular}
\end{table}

\begin{table}[tb]
\caption{The relation between signs ($\pm$) of the MRs and g factors in the $x$-direction magnetic field. }
\label{tab:g-Bx}%
\begin{ruledtabular}
\begin{tabular}{ccc}
$g_{p}^v$ & $g_{p}^c$ & MR$_{x}(B_x)$ \\
\hline
- & -  & -
\\
+ & -   & -
\\
+ & +  & +
\\
-  & + & +
\tabularnewline
\end{tabular}
\end{ruledtabular}
\end{table}

{\color{red}\emph{Roles of orbital moment}}-
The orbital moment has been neglected
in most of the literature
studying the magneto-transport using the semiclassical formalism \cite{Son13prb,Yip15arXiv}.
As shown in Refs.~\cite{Morimoto16prb,Gao17prb},
the orbital moment is essential for the MR anisotropy in a Weyl semimetal.
Moreover, the correction $\bf m\cdot \bf B$ to $\varepsilon(\bf k)$
can enhance the band separation and the negative MR.
To see the effects of the orbital moment, Fig.~\ref{Fig:dependence}
also compares the relative MR in the presence and absence of $\mathbf{m}$.
We can see the orbital moment effectively
{\it enhances} the MR$_x$ a few
times larger. MR$_z$ can be even positive without $\mathbf{m}$.
Therefore, the orbital moment
should be taken into account for quantitatively correct results.

{\color{red}\emph{Discussions}} -
The semiclassical treatment is applicable in the regime where the Landau levels are not well formed. In the quantum limit, where only the lowest band of Landau levels is occupied and MR depends subtly on scattering mechanisms \cite{Lu15prb-QL,Goswami15prb,ZhangSB16njp}, rather than the Berry curvature and orbital moment. Therefore, our ambition is limited in the weak-field limit. The current-jetting effect is usually induced by inhomogeneous currents when attaching point contact electrodes to a large bulk crystal and may also hamper the interpretation of the negative MR data \cite{dosReis16njp}. A recent work by Andreev and Spivak also has pointed out that the negative MR may exist without the chiral anomaly \cite{Andreev17arXiv}.
Equations similar to Eq.~\eqref{Eq:velocity} have been considered from a more general perspective~\cite{Ao07prb}.

\emph{Acknowledgments} - X.D. appreciates the helpful discussions with Hong Yao and Pablo San-Jose. This work was supported by the National Key R \& D Program (Grant No. 2016YFA0301700), National Natural Science Foundation of China (Grant No. 11574127), Guangdong Innovative and Entrepreneurial Research Team Program (Grant No. 2016ZT06D348), and Science, Technology and Innovation Commission of Shenzhen Municipality (Grant No. ZDSYS20170303165926217). X. D. and Z. Z. D. contribute equally to this work.

\bibliographystyle{apsrev4-1-etal-title}
%

\end{document}